\documentstyle[12pt]{article}
\def\SU{\widehat{SU(2)}}
\def\SUm{\widehat{SU(2)}_{\tilde{k}^-}}
\def\SUp{\widehat{SU(2)}_{\tilde{k}^+}}

\def\kp{\tilde{k}^+}
\def\km{\tilde{k}^-}
\def\lp{\ell^+}
\def\lm{\ell^-}
\def\Ap{A^{(+)}}
\def\Am{A^{(-)}}
\def\half{\frac{1}{2}}
\newcommand{\eqn}[1]{(\ref{eq:#1})}

\begin{document}
\begin{titlepage}
\begin{flushright}
DTP/93/47\\
\today \\
\end{flushright}
\vskip 2.cm
\begin{center}
{\Large\bf New identities between unitary minimal Virasoro characters}
\vskip 2cm
{\large Anne Taormina\footnote{email address Anne.Taormina@durham.ac.uk}}
\vskip 0.5cm
{\it Department of Mathematical Sciences, University of Durham, \\
Durham DH1 3LE, England }\\
\vskip 3cm
\end{center}
\begin{center}
{\bf ABSTRACT}
\end{center}
\begin{quote}
Two sets of identities between unitary minimal Virasoro characters
at levels
$m=3,$ 4, 5 are presented and proven.
The first identity suggests a connection between the Ising
and the tricritical Ising models since the $m=3$ Virasoro characters are
obtained as
bilinears of $m=4$ Virasoro characters.
The second identity gives the tricritical
Ising model characters as bilinears in the Ising model characters and the six
combinations of $m=5$ Virasoro characters which do not appear in the spectrum
of the
three state Potts model.
The implication of these identities on the study of the branching rules of
$N=4$
superconformal characters into $\SU \times \SU$ characters is
discussed.
\end{quote}
\vfill
\end{titlepage}

\section{Introduction}
\setcounter{equation}{0}

The theory of unitary highest weight state representations of
$N$-extended superconformal algebras  ($N=0,$ 1, 2, 4) is by now
very well understood
\cite{RC,FQS,GKO,CIZ,BFK,DV1,DV2,NAM,DOB,YU,et1,et2,GUN,pt1,pt2},
and is of considerable interest in the analysis
of the spectrum of string based models.
Unlike the $N=0$, 1, 2 extended superconformal algebras, the $N=4$
algebras both with $\SU$ or $\SU \times \SU$ Kac-Moody subalgebras
have no representations falling in a  minimal series \cite{et1,et2,GUN}.
As a consequence, there is no value of the central charge $c$
allowed by unitarity for which the characters corresponding to
unitary highest weight state representations carry a finite
representation of the modular group.
Generically, there exists a finite number of massless characters
and an infinite tower of massive characters, corresponding to
representations with non-zero and zero Witten index respectively.
The modular transformations of massless characters mix massless
and massive characters in a complicated way,
which has only been analysed in a very limited number of cases.
In the $\SU_k$ extended $N=4$ superconformal algebra, the modular
transformations involve a Mordell integral when the level of $\SU$
is $k=1$ \cite{et3}.
For the $\SUm \times \SUp$ extended $N=4$ algebra, certain combinations
of massless characters play a particular role and transform among
themselves under the modular group when $\kp = \km = 1$ \cite{pt4}.
In attempting to generalise this result for $\kp > 1$ $(\km =1)$,
we have found new relations between unitary minimal Virasoro characters
when $\kp = 2$ and 3.
In the following section, we present these identities, which will be
proven in the Appendix, and we discuss in section 3 their relevance
in the study of $N=4$ superconformal algebras.
Some comments on how the identities between unitary Virasoro characters
at low levels could be generalised to higher levels
are given in the conclusions.
We also briefly discuss the potential consequences of these identities
in 2-d conformal field theory.

\section{The identities}
\setcounter{equation}{0}

The first set of identities relates the unitary Virasoro
characters at levels $m=3$ and $m=4$ in such a way that the three
Ising model characters are given by the vector product of two
3-vectors whose components are the six tricritical Ising model characters,
\begin{equation}
\chi^{Vir~(3)}_{1,i} (q) = \epsilon_{ijk} (-1)^{j+k}
\chi^{Vir~(4)}_{j,4} (q)~\chi^{Vir~(4)}_{k,2} (q).
\label{eq:id1}
\end{equation}
Our definition of the unitary Virasoro
characters at level $m$ is,
\begin{equation}
\chi^{Vir~(m)}_{r,s} (q) = \eta^{-1}(q) \left [
\theta_{r(m+1) -sm,m(m+1)} (q) - \theta_{r(m+1)+sm,m(m+1)}(q) \right],
\label{eq:vir}
\end{equation}
where the integers $m$, $r$ and $s$ have the following ranges,
$$
m=2,3,\ldots;~~~~~~~r=1,2,\ldots,m-1;~~~~~~~s=1,\ldots,r.
$$
The Dedekind function, $\eta(q)$, is defined by,
\begin{equation}
\eta(q) = q^{\frac{1}{24}} \prod_{n=0}^\infty \left (1-q^{n+1}\right),
\end{equation}
while the generalised level $k$ theta functions, $\theta_{m,k}(q)$,
are given by,
\begin{equation}
\theta_{m,k} (q) = \sum_{n \in Z} q^{k\left ( n+\frac{m}{2k}\right )^2},
\label{eq:theta}
\end{equation}
with the properties,
$$
\theta_{m,k}(q) = \theta_{-m,k}(q) = \theta_{2k-m,k}(q).
$$
This definition, \eqn{vir}, of the Virasoro characters coincides with the
definition given in \cite{RC}
up to a factor $q^{-\frac{c}{24}}$ where $c$ is the central charge,
$$
c = 1-\frac{6}{m(m+1)}.
$$
These identities \eqn{id1} can be rewritten in a slightly different
notation where the characters are labelled by their conformal
dimension $h_{r,s}$,
$$
h_{r,s} = \frac{[(m+1)r-ms]^2 -1}{4m(m+1)},
$$
and where the equivalence between
the tricritical Ising model characters and the characters of the
first theory in the superconformal $N=1$ minimal series is implemented.
Since the latter are given by,
\begin{eqnarray}
\chi^{NS}_0(q) & = & \chi^{Vir~(4)}_0(q) + \chi^{Vir~(4)}_{3/2}(q),
\nonumber\\
\tilde\chi^{NS}_0(q) & = & \chi^{Vir~(4)}_0(q) -
\chi^{Vir~(4)}_{3/2}(q),\nonumber\\
\chi^{NS}_{1/10}(q) & = & \chi^{Vir~(4)}_{1/10}(q) +
\chi^{Vir~(4)}_{3/5}(q),\nonumber\\
\tilde\chi^{NS}_{1/10}(q) & = & \chi^{Vir~(4)}_{1/10}(q) -
\chi^{Vir~(4)}_{3/5}(q),\nonumber\\
\chi^{R}_{3/80}(q) & = & \chi^{Vir~(4)}_{3/80}(q),\nonumber\\
\chi^{R}_{7/16}(q) & = & \chi^{Vir~(4)}_{7/16}(q),
\end{eqnarray}
the identities \eqn{id1} take the form,
\begin{eqnarray}
\chi^{Vir~(3)}_{0}(q)+\chi^{Vir~(3)}_{1/2}(q)&=&
\chi^{R}_{3/80}(q)\tilde\chi^{NS}_{0}(q)+
\chi^{R}_{7/16}(q)\tilde\chi^{NS}_{1/10}(q),\nonumber\\
\chi^{Vir~(3)}_{0}(q)-\chi^{Vir~(3)}_{1/2}(q)&=&
\chi^{R}_{3/80}(q)\chi^{NS}_{0}(q)-
\chi^{R}_{7/16}(q)\chi^{NS}_{1/10}(q),\nonumber\\
\chi^{Vir~(3)}_{1/16}(q)&=& \frac{1}{2}\left(
\chi^{NS}_{0}(q)\tilde\chi^{NS}_{1/10}(q)+
\chi^{NS}_{1/10}(q)\tilde\chi^{NS}_{0}(q)\right).
\end{eqnarray}
It is interesting to note that the vector $\chi^{Vir~(3)}_{1,i}(q)$
($i=1,2,3$) being orthogonal to the vectors
$(-1)^j\chi^{Vir~(4)}_{j,4}(q)$
and $(-1)^k\chi^{Vir~(4)}_{k,2}(q)$ (no summation on $j$ and $k$)
is a trivial consequence of the identity \eqn{id1}.
It produces relations which can be derived from repeated use of the
Goddard-Kent-Olive (GKO) sum-rules \cite{GKO},
\begin{equation}
\chi^k_{2\ell}(q,z) \chi^1_{2\ell'}(q,z) = \sum^{k+1}_{2\ell'' = 0}
\chi^{k+1}_{2\ell''} (q,z)
{}~\chi^{Vir~(k+2)}_{2\ell+1,2\ell''+1}(q),
\label{eq:gko}
\end{equation}
where $2\ell'' \equiv 2\ell + 2\ell'~({\rm mod}~2)$ and where the
$SU(2)_k$ characters for isospin $\ell$ ($2\ell = 0,1,\ldots,k$) are
defined by,
\begin{eqnarray}
\chi^{k}_{2\ell} (q,z) &=& q^{-1/8} z^{-1} \prod _{n=1}^\infty (1-q^n)^{-1}
(1-q^nz^2)^{-1}(1-q^{n-1}z^{-2})^{-1}\nonumber \\
&\times &\sum_{m \in Z + \frac{\ell+1/2}{k+2}} q^{(k+2)m^2}
\left[ z^{2(k+2)m}-z^{-2(k+2)m} \right].
\end{eqnarray}
In this instance, one considers the coset $[SU(2)_1 \times SU(2)_1
\times SU(2)_1 ]/SU(2)_3$ and applies the GKO sumrules twice on the
following trilinear in $\SU_1$ characters,
$$
\left [ \chi^1_0 (q,z)~\chi^1_0 (q,z)\right ] \chi^1_1 (q,z) =
 \chi^1_0 (q,z)\left [\chi^1_0 (q,z) ~\chi^1_1 (q,z)\right ].
$$

In the second set of identities, the tricritical Ising model characters
are obtained as the product of unitary Virasoro characters at
levels $m=3$ and $m=5$ in the following way,
\begin{eqnarray}
\chi^{Vir~(4)}_{2,1} (q) & = & \chi^{Vir~(3)}_{2,2}(q)~
\left[ \chi^{Vir~(5)}_{2,1}(q)-\chi^{Vir~(5)}_{3,1}(q)\right], \nonumber\\
\chi^{Vir~(4)}_{2,2} (q) & = & \chi^{Vir~(3)}_{2,2}(q)~
\left[ \chi^{Vir~(5)}_{1,1}(q)-\chi^{Vir~(5)}_{4,1}(q)\right], \nonumber\\
\chi^{Vir~(4)}_{1,1} (q)\pm \chi^{Vir~(4)}_{3,1} (q) & = &
\left[\chi^{Vir~(3)}_{1,1}(q)\pm \chi^{Vir~(3)}_{2,1}(q)\right]~
\left[ \chi^{Vir~(5)}_{2,2}(q)\mp\chi^{Vir~(5)}_{3,2}(q)\right], \nonumber\\
\chi^{Vir~(4)}_{1,2} (q)\pm \chi^{Vir~(4)}_{3,2} (q) & = &
\left[\chi^{Vir~(3)}_{1,1}(q)\pm \chi^{Vir~(3)}_{2,1}(q)\right]~
\left[ \chi^{Vir~(5)}_{1,2}(q)\mp\chi^{Vir~(5)}_{4,2}(q)\right].
\label{eq:id2}
\end{eqnarray}
It is remarkable that the six combinations of level $m=5$ Virasoro
characters involved are precisely those which do {\em not} appear
in the spectrum of the three state Potts model.
The identities \eqn{id2} are consistent with the weaker identities,
\begin{equation}
\chi^{Vir~(4)}_{2,2}(q)~
\left[ \chi^{Vir~(5)}_{2,1}(q)-\chi^{Vir~(5)}_{3,1}(q)\right]
=
\chi^{Vir~(4)}_{2,1}(q)~
\left[ \chi^{Vir~(5)}_{1,1}(q)-\chi^{Vir~(5)}_{4,1}(q)\right],
\end{equation}
and,
\begin{eqnarray}
\lefteqn{\left[\chi^{Vir~(4)}_{1,1}(q)\pm \chi^{Vir~(4)}_{3,1}(q)\right] ~
\left[ \chi^{Vir~(5)}_{1,2}(q)\mp\chi^{Vir~(5)}_{4,2}(q)\right]}\nonumber \\
&=&\left[\chi^{Vir~(4)}_{1,2}(q)\pm \chi^{Vir~(4)}_{3,2}(q)\right]~
\left[ \chi^{Vir~(5)}_{2,2}(q)\mp\chi^{Vir~(5)}_{3,2}(q)\right],
\end{eqnarray}
which can be obtained from the GKO character sumrules for the coset
$[SU(2)_1 \times SU(2)_2 \times SU(2)_1 ]/SU(2)_4$ when considering the
following trilinears in  $\SU$ characters,
$$
\left [ \chi^1_0 (q,z)~\chi^2_0 (q,z)\right ] \chi^1_1 (q,z) =
 \chi^1_0 (q,z)\left [\chi^2_0 (q,z) ~\chi^1_1 (q,z)\right ],
$$
and,
$$
\left [ \chi^1_0 (q,z)~\chi^2_1 (q,z)\right ] \chi^1_1 (q,z) =
 \chi^1_0 (q,z)\left [\chi^2_1 (q,z) ~\chi^1_1 (q,z)\right ].
$$
A proof of the identities \eqn{id1} and \eqn{id2} involving the Jacobi
triple product identity and standard properties of the generalised theta
functions \eqn{theta} is given in the Appendix.

In order to gain some insight in the way one might generalise this
type of relations between unitary Virasoro characters,
we now turn to  $N=4$ superconformal algebras whose study has prompted
the identities  presented here.

\section{Properties of $N=4$ superconformal characters}
\setcounter{equation}{0}

The $\SUp \times \SUm$ extended $N=4$ algebra we consider is a
non-linear superconformal algebra which, together with a dimension
2 Virasoro generator $L(z)$ and 4 dimension $3/2$ supercurrents
$G^a(z)$ $(a=1,2,3,4)$, contains 6 currents $T^{\pm i}(z)$ ($i=1,2,3$)
which are dimension 1 primaries with respect to $L(z)$ and generate
two Kac-Moody $\SU$ algebras at levels $\kp$ and $\km$ \cite{Sevrin}.
The representation theory and corresponding characters were given in
\cite{GUN,pt1,pt2}
for unitary highest weight state representations.
These are labelled by the two isospin quantum numbers $\ell^+$ and
$\ell^-$ and have conformal dimension $\tilde{h}$ whose lower bound
$\tilde{h}_0$ is a function of $\ell^+$, $\ell^-$, $\kp$ and $\km$.
An irreducible representation with conformal dimension $\tilde{h}_0$
is called massless or chiral (in the Ramond sector, it corresponds
to a representation with non-zero Witten index), while any
representation with conformal dimension $\tilde{h}> \tilde{h}^0$
is called massive.
For fixed $\kp$, $\km$, there is a finite number of massless and
an infinite
number of massive representations and corresponding characters.
In the Ramond sector, the massless characters are labelled by,
$$
Ch_0^R (\kp, \km, \lp, \lm, \tilde{h}^R_0 ; q, z_+, z_- ),
$$
with $\ell^\pm = 0,1/2,\ldots, \tilde{k}^\pm/2$ and,
$$
\tilde{h}^R_0 = \frac{\kp\km + 4(\lp+\lm) (\lp+\lm + 1)}{4(\kp+\km+2)},
$$
while massive characters are denoted by,
$$
Ch_m^R (\kp, \km, \lp, \lm, \tilde{h}^R ; q, z_+, z_- ),
$$
with $\ell^\pm = 1/2,\ldots, \tilde{k}^\pm/2$ and $\tilde{h}^R >
\tilde{h}^R_0$.

Two essential properties of the characters are used in the following.
First, the massive characters split into two massless ones as $\tilde{h}^R$
reaches the lower bound $\tilde{h}^R_0$,
\begin{eqnarray}
\lefteqn{Ch_m^R (\kp, \km, \lp, \lm+\half, \tilde{h}^R ; q, z_+, z_- )
= q^{\tilde{h}^R-\tilde{h}^R_0}  }
\nonumber \\
&\times &\left [
Ch_0^R (\kp, \km, \lp, \lm, \tilde{h}^R_0 ; q, z_+, z_- )
+
Ch_0^R (\kp, \km, \lp-\half, \lm+\half, \tilde{h}^R_0 ; q, z_+, z_- ) \right ].
\nonumber \\
\label{eq:chm}
\end{eqnarray}
Furthermore, when the conformal dimension is,
$$
\tilde{h}^R_m \equiv \frac{(\kp+1) (\km +1)}{4(\kp+\km +2)}
\left[ \frac{2\lp}{\kp+1} - \frac{2\lm}{\km+1} \right ] ^2
+\frac{\left[(2\lp+2\lm)^2+\kp\km-1 \right]}{4(\kp+\km +2)} ,
$$
one has,
\begin{eqnarray}
\lefteqn{\eta(q)~Ch_m^R (\kp, \km, \lp, \lm, \tilde{h}^R_m ; q, z_+, z_- )
}\nonumber \\
&=&
\sum_{2\lambda^+ = 0}^{\kp}
\sum_{2\lambda^- = 0}^{\km}
\chi^{\kp}_{2\lambda^+} (q, z_+)
{}~\chi^{\km}_{2\lambda^-} (q, z_-)
{}~\chi^{Vir~(\kp+1)}_{2\lp,2\lambda^++1} (q)
{}~\chi^{Vir~(\km+1)}_{\km-2\lm+1,\km-2\lambda^-+1} (q),
\label{eq:prop1}
\end{eqnarray}
where $2\lambda^+ -  2\lambda^- \equiv 2\lp-2\lm +1 ~({\rm mod}~2)$,
so that the branching functions for {\em massive} characters into
$\SUp \times \SUm$ characters $\chi^{\kp}_{2\lambda^+} (q, z_+)
{}~\chi^{\km}_{2\lambda^-} (q, z_-)$ are products of unitary
Virasoro characters at levels $m=\kp+1$ and $m = \km +1$.
For $\tilde{k}^\pm = 1$, we recall that
$\chi^{Vir~(\tilde{k}^\pm+1)}_{2\ell,2\ell'}(q) = 1$.

The second property of interest is that when the angular variables
$z_\pm$ are related by $z_- = - z_+^{-1}$, the massive characters
in the Ramond sector vanish while the massless characters reduce
to $\SU_{\kp+\km}$ characters \cite{opt},
\begin{eqnarray}
Ch_m^R (\kp, \km, \lp, \lm, \tilde{h}^R_m ; q, z_+, - z_+^{-1} )&=&0,
\nonumber \\
Ch_0^R (\kp, \km, \lp, \lm, \tilde{h}^R_0 ; q, z_+, - z_+^{-1} )&=&
(-1)^{2\lm} \chi_{2(\lp+\lm)}^{\kp+\km} (q,z_+).
\label{eq:prop2}
\end{eqnarray}

Let us now introduce the following combinations of massless characters
for $\km = 1$, which is the only case discussed here (for simplicity
the $\kp$, $\km$ and $\tilde{h}^R_0$ arguments have been suppressed),
\begin{eqnarray}
Ch_0^R (L=0; q, z_+, z_-) &\equiv & -Ch_0^R(\lp = 0,\lm = 0; q, z_+, z_-),
\nonumber \\
Ch_0^R (L=1,\ldots,\kp; q, z_+, z_-) &\equiv & \half
\left[ Ch_0^R\left(\half (L-1),\half ; q, z_+, z_-\right ) -
Ch_0^R\left(\half L,0 ; q, z_+, z_-\right ) \right ],\nonumber \\
Ch_0^R (L=\kp+1; q, z_+, z_-) &\equiv &
Ch_0^R\left(\half \kp, \half ; q, z_+, z_-\right ),
\label{eq:combo}
\end{eqnarray}
where $ 2(\lp+\lm) = L$.
The combinations for $L = 1,\ldots, \kp$
are orthogonal to the combinations present in \eqn{chm}.

In terms of $\SUp \times \SU_1$ characters, the above combinations are
given by,
\begin{equation}
Ch_0^R(L;q, z_+, z_-) = \sum_{2\lambda^+ = 0}^{\kp} \sum _{2\lambda^- =0}^1
(-1)^{2\lambda^++L+1} ~Y_{2\lambda^++1, L+1;2\lambda^-}^{(\kp+2)} (q)
{}~\chi_{2\lambda^+}^{\kp} (q,z_+)
{}~\chi_{2\lambda^-}^1 (q,z_-),
\label{eq:ch0br}
\end{equation}
where $L \equiv 2\lambda^++2\lambda^-$ mod 2 and the branching functions
$Y_{2\lambda^++1, L+1;2\lambda^-}^{\kp+2} (q)$ are not
known for $\kp > 1$.
It turns out that,
\begin{equation}
Y_{2\lambda^++1, L+1;2\lambda^-}^{(\kp+2)} (q)=
Y_{\kp+1-2\lambda^+, \kp+2-L;1-2\lambda^-}^{(\kp+2)} (q),
\label{eq:Ysym}
\end{equation}
so that there are as many such functions as unitary Virasoro
characters at level $m=\kp+2$.
We find it convenient to introduce the following notation ($\km =1$),
\begin{equation}
Y_{2\lambda^++1, L+1;2\lambda^-}^{(\kp+2)} (q)=
q^{\frac{(2\lambda^+-L)^2-(2\lambda^-)^2}{4}}
{}~~Y_{2\lambda^++1, L+1}^{(\kp+2)} (q),
\end{equation}
where the functions $Y_{2\lambda^++1, L+1}^{(\kp+2)} (q)$
have a $q$ expansion,
$$
Y_{2\lambda^++1, L+1}^{(\kp+2)} (q) \sim
q^{\frac{c}{24} -h_{2\lambda^++1, L+1} }~\sum_{n=0}^\infty a_n q^n,
{}~~~~~~~a_n \in Z,
$$
characterised by a prefactor which is the inverse of the
corresponding $m=\kp+2$ Virasoro character prefactor.
We emphasize that the symmetry property \eqn{Ysym} does not necessarily
imply a similar symmetry property on the functions
$Y_{2\lambda^++1, L+1}^{(\kp+2)} (q)$. For instance, when $\kp = 1$,
$Y_{2,1}^{(3)} (q) = q~Y_{1,3}^{(3)}(q)$.

When $\kp=1$, it is easy to calculate the three branching functions
\cite{opt,pt4},
which can be written in terms of Ising model characters,
\begin{eqnarray}
Y_{1,1;0}^{(3)} (q) \pm Y_{2,1;1}^{(3)} (q)
=Y_{1,1}^{(3)} (q) \pm Y_{2,1}^{(3)} (q) &=&
\frac{1}{\chi^{Vir~(3)}_{1,1}(q)
\pm\chi^{Vir~(3)}_{2,1}(q)},\nonumber \\
Y_{2,2;0}^{(3)} (q) =Y_{2,2}^{(3)} (q)  &=&
\frac{1}{2\chi^{Vir~(3)}_{2,2}(q) }.
\label{eq:branch}
\end{eqnarray}

In an attempt to determine the branching functions
for $\kp > 1$ \cite{pt5},
we analyse the
properties of the matrix $\Ap$ of branching functions for the
combinations \eqn{combo} of massless characters $Ch_0^R(2L)$,
($L=0,\ldots, [\half (\kp+1)]$)
and the massive characters $\eta(q) Ch_m^R(L,\half)$
($L=1,\ldots, [\half \kp]$) as well as the properties of the
matrix $\Am$
of branching functions for the combinations of massless characters
$Ch_0^R(2L+1)$
($L=0,\ldots, [\half \kp]$) and massive characters
$\eta(q) Ch_m^R(L+\half,\half)$ ($L=0,\ldots,
[\half (\kp-1)]$).  By $[r]$, we mean as usual the integer part
of the real number $r$, i.e. the biggest integer smaller than or
equal to $r$.
The matrices $\Ap$ and $\Am$ are both of dimension $(\kp+1) \times (\kp+1)$ ,
and have the same determinant (up to a sign) when $\kp$ is even,
$$
\det \Ap = - \det \Am ~~~~~~~{\rm for}~~~\kp \in 2 Z.
$$
The inverse matrices, $\left [\Ap\right ]^{-1} $ and
$\left [\Am\right ]^{-1} $,
are encoded in the relation,
\begin{eqnarray}
\chi_{2\lp}^{\kp} (q,z_+)
{}~\chi_{2\lm}^1 (q,z_-) &=& (-1)^{2\lm +1}
\sum_{\stackrel{2\ell =0}{2\ell
\equiv 2\lp +2\lm \ \mbox{mod}\  2}}^{\kp+1} Ch_0^R(2\ell ; q,z_+,z_-)
{}~\chi^{Vir~(\kp+2)}_{2\lp+1,2\ell+1} (q)\nonumber \\
&+& \eta(q)
\sum_{\stackrel{2\ell =1}{2\ell
\equiv 2\lp +2\lm \ \mbox{mod}\  2}}^{\kp}
Ch_m^R\left (\ell,\half,\tilde h^R_m; q,z_+,z_-\right )
{}~X^{(\kp+1)}_{2\ell,2\lp+1;2\lm} (q),\nonumber \\
\label{eq:inv}
\end{eqnarray}
for $ L = 2(\lp+\lm)$  being even and odd respectively.
The functions $X^{(\kp+1)}_{2\ell,2\lp+1;2\lm} (q)$ are unknown, except
for $\kp = 1$ where they are equal to 1.
They obey the symmetry,
\begin{equation}
X^{(\kp+1)}_{2\ell,2\lp+1;2\lm} (q) =
X^{(\kp+1)}_{\kp+1-2\ell,\kp+1-2\lp;1-2\lm} (q),
\end{equation}
and there are as many such functions as Virasoro characters
at level $m = \kp+1$.
(Recall that when $\km = 1$, $2\lm = 0$ or 1 when $2\ell -2\lp$ is
even or odd respectively.)
We define,
\begin{equation}
X^{(\kp+1)}_{2\ell,2\lp+1;2\lm} (q) =
q^{(\ell-\lp)(\ell-\lp-1)+(\lm)^2}~X^{(\kp+1)}_{2\ell,2\lp+1} (q),
\end{equation}
where the $q$ expansion of the functions $X^{(\kp+1)}_{2\ell,2\lp+1} (q)$
is,
$$
X^{(\kp+1)}_{2\ell,2\lp+1} (q) = q^{\frac{c}{24}-h_{2\ell,2\lp+1}}~
\sum_{n=0}^\infty~b_n q^n,~~~~~~~~b_n \in Z,
$$
with the prefactor being the inverse
of the corresponding level $m=\kp+1$ Virasoro character.

Note that this relation \eqn{inv} reduces to the GKO sumrule when
$z_- = -z_+^{-1}$, as can be easily seen using the properties \eqn{prop2} and
$$
\chi^k_{2\ell} (q, -z^{-1}) = (-1)^{2\ell} \chi^k_{2\ell} (q,z).
$$
The key observation which led us to the identities, \eqn{id1} and \eqn{id2},
presented in the previous section is that the product of the determinants
of the matrices $\Ap$ and $\Am$ is equal to minus one when $\kp = 1$.
Indeed, the matrices $\Ap$ and $\Am$ in this case are given by
(suppressing the $q$ dependence),
\begin{equation}
\Ap = \left (
\begin{array}{cc}
-Y^{(3)}_{1,1;0}  & Y^{(3)}_{2,1;1} \\
-Y^{(3)}_{2,1;1}   & Y^{(3)}_{1,1;0}
\end{array}
\right ),~~~~~~
\Am = \left (
\begin{array}{cc}
Y^{(3)}_{2,2;0}   & -Y^{(3)}_{2,2;0} \\
1                &  1
\end{array}
\right ),
\end{equation}
and,
\begin{equation}
\det \Ap \det \Am =  \frac{1}{\chi^{Vir~(3)}_{2,2} (q)
{}~\left (\left [\chi^{Vir~(3)}_{2,1} (q)\right ]^2
-\left [\chi^{Vir~(3)}_{1,1} (q)\right ]^2\right )} = -1.
\end{equation}
This can be seen by using the expressions  \eqn{branch} and the well known
identity,
$$
1 = \chi^{Vir~(3)}_{2,2} (q)
{}~\left(\chi^{Vir~(3)}_{1,1} (q) +\chi^{Vir~(3)}_{2,1} (q)\right)
{}~\left(\chi^{Vir~(3)}_{1,1} (q) -\chi^{Vir~(3)}_{2,1} (q)\right),
$$
which can be easily checked by using the infinite product
representation of Ising model
characters given in the Appendix.
There is some evidence that the product $\det \Ap \det \Am$ is
a modular invariant \cite{jens}
and we therefore conjecture the result,
\begin{equation}
\det \Ap \det \Am = -1, ~~~\forall ~\kp \in N, ~\km = 1.
\label{eq:conj1}
\end{equation}
The identities of section 2 enable us to prove it for $\kp = 2$.

Indeed for $\kp =2$, we have, according to the relations \eqn{ch0br}
and \eqn{inv},
\begin{eqnarray}
\Ap &=&
\left (
\begin{array}{ccc}
      -Y^{(4)}_{1,4;1}  &        Y^{(4)}_{2,4;0}  & -Y^{(4)}_{3,4;1} \\
      -Y^{(4)}_{1,2;1}  &        Y^{(4)}_{2,2;0}  & -Y^{(4)}_{3,2;1} \\
\chi^{Vir~(3)}_{1,1}  & \chi^{Vir~(3)}_{1,2}  & \chi^{Vir~(3)}_{1,3}
\end{array}
\right ),\nonumber \\
\left [ \Ap \right ] ^{-1}  &=&
\left (
\begin{array}{ccc}
-\chi^{Vir~(4)}_{1,4}  & -\chi^{Vir~(4)}_{1,2}  & X^{(3)}_{1,1;1} \\
 \chi^{Vir~(4)}_{2,4}  &  \chi^{Vir~(4)}_{2,2}  & X^{(3)}_{1,2;0} \\
-\chi^{Vir~(4)}_{3,4}  & -\chi^{Vir~(4)}_{3,2}  & X^{(3)}_{1,3;1}
\end{array}
\right ),\nonumber \\
\Am  &=&
\left (
\begin{array}{ccc}
       Y^{(4)}_{1,2;1}  &       -Y^{(4)}_{2,2;0}  & Y^{(4)}_{3,2;1} \\
       Y^{(4)}_{1,4;1}  &       -Y^{(4)}_{2,4;0}  & Y^{(4)}_{3,4;1} \\
\chi^{Vir~(3)}_{1,1}  & \chi^{Vir~(3)}_{1,2}  & \chi^{Vir~(3)}_{1,3}
\end{array}
\right ),\nonumber \\
\left [ \Am \right ] ^{-1}  &=&
\left (
\begin{array}{ccc}
 \chi^{Vir~(4)}_{1,2}  &  \chi^{Vir~(4)}_{1,4}   & X^{(3)}_{1,1;1} \\
-\chi^{Vir~(4)}_{2,2}  & -\chi^{Vir~(4)}_{2,4}   & X^{(3)}_{1,2;0} \\
 \chi^{Vir~(4)}_{3,2}  &  \chi^{Vir~(4)}_{3,4}   & X^{(3)}_{1,3;1}
\end{array}
\right ).
\end{eqnarray}
Exploiting the fact that the matrices $\Ap$ and $\left [\Ap\right ]^{-1} $
are inverses
of each other, we may write,
\begin{equation}
\chi^{Vir~(3)}_{1,i} (q) = \frac{\epsilon_{ijk} (-1)^{j+k}
{}~\chi^{Vir~(4)}_{j,4}(q)~\chi^{Vir~(4)}_{k,2}(q)}{\det \left [\Ap\right
]^{-1} },
\end{equation}
where,
$$
\det \left [\Ap\right ]^{-1}  = \epsilon_{abc} (-1)^{a+b}
{}~\chi^{Vir~(4)}_{a,4}(q)
{}~\chi^{Vir~(4)}_{b,2}(q)~X^{(3)}_{1,c;\sigma} (q)
$$
with $\sigma = 0$ or 1 for $1-c$ odd or even respectively.
By virtue of \eqn{id1},
$$
\det \left [\Ap\right ]^{-1} =1.
$$
The result \eqn{conj1} follows from the
observation that $\det \Am = -\det \Ap$ when $\kp =2$.

When $\kp=3$, the situation is slightly different since the matrices
$\Ap$ and $\Am$ do not have the same determinant up to a sign.
With the matrices
$\Ap$ and $\left [\Ap\right ]^{-1} $ taken to be,
\begin{eqnarray}
\Ap &=&
\left (
\begin{array}{cccc}
Y^{(5)}_{1,5;0}  & -Y^{(5)}_{2,5;1}  & Y^{(5)}_{3,5;0} &-Y^{(5)}_{4,5;1}  \\
Y^{(5)}_{1,3;0}  & -Y^{(5)}_{2,3;1}  & Y^{(5)}_{3,3;0} &-Y^{(5)}_{4,3;1}  \\
Y^{(5)}_{1,1;0}  & -Y^{(5)}_{2,1;1}  & Y^{(5)}_{3,1;0} &-Y^{(5)}_{4,1;1}  \\
\chi^{Vir~(4)}_{2,1}  & \chi^{Vir~(4)}_{2,2}  & \chi^{Vir~(4)}_{2,3}
& \chi^{Vir~(4)}_{2,4}
\end{array}
\right ),\nonumber \\
&&\nonumber \\
\left [ \Ap \right ] ^{-1}  &=&
\left (
\begin{array}{cccc}
 \chi^{Vir~(5)}_{1,5}  &  \chi^{Vir~(5)}_{1,3}  &  \chi^{Vir~(5)}_{1,1}
&  X^{(4)}_{2,1;0} \\
-\chi^{Vir~(5)}_{2,5}  & -\chi^{Vir~(5)}_{2,3}  & -\chi^{Vir~(5)}_{2,1}
 &  X^{(4)}_{2,2;1} \\
 \chi^{Vir~(5)}_{3,5}  &  \chi^{Vir~(5)}_{3,3}  &  \chi^{Vir~(5)}_{3,1}
&  X^{(4)}_{2,3;0} \\
-\chi^{Vir~(5)}_{4,5}  & -\chi^{Vir~(5)}_{4,3}  & -\chi^{Vir~(5)}_{4,1}
&  X^{(4)}_{2,4;1}
\end{array}
\right ),
\end{eqnarray}
it is easy to see that,
\begin{equation}
\chi^{Vir~(4)}_{2,i} (q) = \frac{\epsilon_{ijkl} (-1)^{j+k+l}
{}~\chi^{Vir~(5)}_{j,5}(q)~\chi^{Vir~(5)}_{k,3}(q)~\chi^{Vir~(5)}_{l,1}(q)}
{\det \left [\Ap\right ]^{-1} }.
\label{eq:rel45}
\end{equation}
However, because the following bilinear in $m=5$ Virasoro characters
is equal to 1 (see Appendix),
\begin{equation}
\epsilon_{ab} \left ( \chi^{Vir~(5)}_{a,1} (q)
+ \chi^{Vir~(5)}_{a,5} (q) \right ) \chi^{Vir~(5)}_{b,3} (q) = 1,
\label{eq:d5}
\end{equation}
one obtains from \eqn{rel45} that,
\begin{eqnarray}
\chi^{Vir~(4)}_{2,1} (q) &=&
\frac{\chi^{Vir~(5)}_{2,1}(q)-\chi^{Vir~(5)}_{2,5}(q)}
{\det \left [\Ap\right ]^{-1} },\nonumber \\
\chi^{Vir~(4)}_{2,2} (q) &=&
\frac{\chi^{Vir~(5)}_{1,1}(q)-\chi^{Vir~(5)}_{1,5}(q)}
{\det \left [\Ap\right ]^{-1} }.
\label{eq:4to5}
\end{eqnarray}
We have used the well known symmetry properties of unitary Virasoro characters,
$$
\chi^{Vir~(m)}_{r,s} (q) = \chi^{Vir~(m)}_{m-r,m+1-s} (q).
$$
It is now straightforward to conclude from the identities \eqn{id2} and the
expression \eqn{4to5} that
$$
\det \left [\Ap\right ]^{-1}  = \frac{1}{\chi^{Vir~(3)}_{2,2} (q)}.
$$
The number of unknown functions, $X^{(4)}_{r,s;\sigma }(q)$,  in the
matrix $\left [\Am\right ]^{-1} $
is however too high to
derive its determinant, even with the help of the identities
\eqn{id2}.
This precisely shows the limit of our approach to generate more
identities of the type described in section 2.
Indeed, the higher value $\kp$ takes, the higher number of unknown
branching functions
$Y^{(\kp+2)}_{r,s;\sigma}(q)$ and functions $X^{(\kp+1)}_{r,s;\sigma}(q)$
one gets.
For instance, when $\kp = 4$, the fact that $\Ap$ and
$\left [\Ap\right ]^{-1} $ are inverse matrices implies that the
$m=5$ Virasoro characters are obtained as,
\begin{eqnarray}
\chi^{Vir~(5)}_{1,i}(q) &=&
\frac{1}{\det \left [\Ap\right ]^{-1} }~\epsilon_{ijklm} (-1)^{j+k+l}
{}~\chi^{Vir~(6)}_{j,6}(q)~\chi^{Vir~(6)}_{k,4}(q)~\chi^{Vir~(6)}_{l,2}(q)~
X^{(5)}_{3,m;\sigma}(q),\nonumber \\
\chi^{Vir~(5)}_{3,i}(q) &=& \frac{1}{\det \left [\Ap\right ]^{-1} }
{}~\epsilon_{ijklm} (-1)^{j+k+l} ~\chi^{Vir~(6)}_{j,6}(q)
{}~\chi^{Vir~(6)}_{k,4}(q)~\chi^{Vir~(6)}_{l,2}(q)~
X^{(5)}_{1,m;\sigma}(q),\nonumber \\
\end{eqnarray}
so that potential identities between unitary Virasoro characters
depend on the unknown functions
$X^{(5)}_{r,s;\sigma}(q)$.

\newpage
\section{Conclusions}
\setcounter{equation}{0}

We have presented and proven two sets of identities between unitary minimal
Virasoro characters at levels $m=3$, 4, 5.
The first identity \eqn{id1} suggests a strong connection between the Ising
and the tricritical Ising models since the $m=3$ Virasoro characters are
obtained as
bilinears of $m=4$ Virasoro characters.
It may also imply an as yet unknown mechanism which produces a conformal field
theory with central charge $c=\half$
when considering two copies of a $c=\frac{7}{10}$ theory in a much less trivial
way than their tensor product.
Such a mechanism would be a very interesting alternative to twisting the energy
momentum tensor of a given theory in order to alter its central charge.

The second identity \eqn{id2} is more involved since it gives the tricritical
Ising model characters as bilinears in the Ising model characters and the six
combinations of $m=5$ Virasoro characters which do not appear in the spectrum
of the  three state Potts model.
A field theoretic interpretation of these identities would certainly shed new
light on the underlying structure of minimal Virasoro theories.

It would also be important to investigate the generalisation of these
identities to higher level Virasoro characters.  Our approach, which
involves the study of $N=4$ superconformal characters and their branching
functions into $SU(2) \times SU(2)$ characters is quite limited at present
due to the lack of information on the  analytic structure of the branching
functions.
A more direct approach based on a deeper field theoretic understanding
would undoubtedly reveal more relations between Virasoro characters of
different levels.

\section*{Acknowledgements}

We thank Jens-Lyng Petersen
for sharing his insights into the extended $N=4$
superconformal algebra.
We also thank Ed Corrigan, Paul Mansfield and
Gerard Watts for stimulating discussions.
We acknowledge the U.K. Science
and Engineering Research Council for the award of an Advanced Fellowship.

\newpage
\appendix
\section{Appendix: Proof of the identities}
\setcounter{equation}{0}

We give here a complete proof of the vectorial identity \eqn{id1}
between $m=3$ and $m=4$ unitary Virasoro characters,
as well as a proof  of one of the identities \eqn{id2} between
$m=3,4,5$ Virasoro characters.  The other identities in this second
set can be proven by similar methods.
We use the infinite product representation of $m=3$ and $m=4$ characters,
given for instance in \cite{RC},
\begin{eqnarray}
\chi^{Vir~(3)}_{1,2} (q) & = & q^{\frac{1}{24}}
\prod_{n=0}^\infty \left(1+q^{n+1}\right),\nonumber \\
\chi^{Vir~(3)}_{1,1}(q) \pm \chi^{Vir~(3)}_{1,3} (q) & = & q^{-\frac{1}{48}}
\prod_{n=0}^\infty \left(1\pm q^{n+\half}\right),
\label{eq:chi3}
\end{eqnarray}
and,
\begin{eqnarray}
\chi^{Vir~(4)}_{2,1} (q) & = & q^{-\frac{7}{240}+\frac{7}{16}}
\prod_{n=0}^\infty \frac{\left(1+q^{5n+1}\right)
{}~\left(1+q^{5n+4}\right)~\left(1+q^{5n+5}\right)}
{\left(1-q^{5n+2}\right)~\left(1-q^{5n+3}\right)}
,\nonumber \\
\chi^{Vir~(4)}_{2,2} (q) & = & q^{-\frac{7}{240}+\frac{3}{80}}
\prod_{n=0}^\infty \frac{\left(1+q^{5n+2}\right)
{}~\left(1+q^{5n+3}\right)~\left(1+q^{5n+5}\right)}
{\left(1-q^{5n+1}\right)~\left(1-q^{5n+4}\right)}
,\nonumber \\
\chi^{Vir~(4)}_{1,1} (q) \pm \chi^{Vir~(4)}_{3,1} (q)& = & q^{-\frac{7}{240}}
\prod_{n=0}^\infty \frac{\left(1\pm q^{5n+\frac{3}{2}}\right)
{}~\left(1\pm q^{5n+\frac{5}{2}}\right)~\left(1\pm q^{5n+\frac{7}{2}}\right)}
{\left(1-q^{5n+2}\right)~\left(1-q^{5n+3}\right)}
,\nonumber \\
\chi^{Vir~(4)}_{1,2} (q) \pm \chi^{Vir~(4)}_{3,2} (q)& = &
q^{-\frac{7}{240}+\frac{1}{10}}
\prod_{n=0}^\infty \frac{\left(1\pm q^{5n+\frac{1}{2}}\right)
{}~\left(1\pm q^{5n+\frac{5}{2}}\right)~\left(1\pm q^{5n+ \frac{9}{2}}\right)}
{\left(1-q^{5n+1}\right)~\left(1-q^{5n+4}\right)}.
\label{eq:chi4}
\end{eqnarray}

The vectorial identity \eqn{id1} is equivalent to the set of three scalar
identities,
\begin{eqnarray}
\chi^{Vir~(3)}_{1,2} (q) & = & \half \Biggl (
 \left [ \chi^{Vir~(4)}_{1,1} (q) + \chi^{Vir~(4)}_{3,1} (q) \right ]
{}~\left [ \chi^{Vir~(4)}_{1,2}(q)  - \chi^{Vir~(4)}_{3,2} (q) \right ]
\nonumber \\
&&+\left [ \chi^{Vir~(4)}_{1,1}(q)  - \chi^{Vir~(4)}_{3,1}(q)  \right ]
{}~\left [ \chi^{Vir~(4)}_{1,2}(q)  + \chi^{Vir~(4)}_{3,2}(q)  \right ]
\Biggr ),\nonumber \\
&& \label{eq:a3} \\
\chi^{Vir~(3)}_{1,1}(q)\pm \chi^{Vir~(3)}_{1,3}(q)  & = &
\chi^{Vir~(4)}_{2,2}(q)
{}~ \left [ \chi^{Vir~(4)}_{1,1} (q) \mp \chi^{Vir~(4)}_{3,1} (q) \right
]\nonumber \\
&&\pm \chi^{Vir~(4)}_{2,1}(q)
{}~ \left [ \chi^{Vir~(4)}_{1,2} (q) \mp \chi^{Vir~(4)}_{3,2}  (q)\right ].
\label{eq:a4}
\end{eqnarray}
When using the product representation \eqn{chi3} and \eqn{chi4},
equations \eqn{a3} and \eqn{a4} may be rewritten
as,
\begin{eqnarray}
2\prod_{n=0}^\infty \frac{\left(1+q^{n+1}\right)~\left(1-q^{n+1}\right)}
{\left(1-q^{10n+5}\right)~\left(1-q^{5n+5}\right)}
&=&
\prod_{n=0}^\infty \left(1+q^{5n+\frac{3}{2}}\right)
{}~\left(1+q^{5n+\frac{7}{2}}\right)
{}~\left(1-q^{5n+\frac{1}{2}}\right)~\left(1-q^{5n+\frac{9}{2}}\right)~
\nonumber \\
&+&\prod_{n=0}^\infty \left(1-q^{5n+\frac{3}{2}}\right)
{}~\left(1-q^{5n+\frac{7}{2}}\right)
{}~\left(1+q^{5n+\frac{1}{2}}\right)~\left(1+q^{5n+\frac{9}{2}}\right),
\nonumber \\
\label{eq:a5}
\end{eqnarray}
and,
\begin{eqnarray}
\lefteqn{\prod_{n=0}^\infty \frac{\left(1\pm q^{n+\half}\right)
{}~\left(1-q^{n+1}\right)}
{\left(1-q^{5n+5}\right)~\left(1+q^{5n+5}\right)
{}~\left(1\mp q^{5n+\frac{5}{2}}\right)}
}
\nonumber \\
&=&\prod_{n=0}^\infty \left(1+q^{5n+2}\right)~\left(1+q^{5n+3}\right)
{}~\left(1\mp q^{5n+\frac{3}{2}}\right)~\left(1\mp q^{5n+\frac{7}{2}}\right)~
\nonumber \\
&\pm &q^{\half}\prod_{n=0}^\infty \left(1+q^{5n+1}\right)
{}~\left(1+q^{5n+4}\right)~\left(1\mp q^{5n+\frac{1}{2}}\right)
{}~\left(1\mp q^{5n+\frac{9}{2}}\right).
\label{eq:a6}
\end{eqnarray}

The Jacobi triple product identity,
\begin{equation}
\sum_{n=-\infty}^{+\infty} z^n q^{n^2} = \prod_{n=0}^\infty
\left(1-q^{2n+2}\right)~\left(1+z
q^{2n+1}\right)~\left(1+z^{-1}q^{2n+1}\right),
\end{equation}
with $q \to q^{\frac{5}{2}}$, $z \to \pm q^{\frac{5}{2}-i}$ for
$i = \half$, $\frac{3}{2}$, $1$ and $3$,
allows us to rewrite \eqn{a5} and \eqn{a6} as,
\begin{eqnarray}
\lefteqn{2\prod_{n=0}^\infty \frac{\left(1+q^{n+1}\right)
{}~\left(1-q^{n+1}\right)~\left(1-q^{5n+5}\right)}
{\left(1-q^{10n+5}\right)}}
\nonumber \\
&=&
\left [ \sum_{n=-\infty}^{+\infty} q^{\frac{5}{2}n^2 +n} \right ]
{}~\left [ \sum_{m=-\infty}^{+\infty} (-1)^m q^{\frac{5}{2}m^2 +2m} \right ]
+\left [ \sum_{n=-\infty}^{+\infty} q^{\frac{5}{2}n^2 +2n} \right ]
{}~\left [ \sum_{m=-\infty}^{+\infty} (-1)^m q^{\frac{5}{2}m^2 +m} \right ]
\nonumber \\
&=& 2 q^{-\half} \Biggl [
\theta_{2,10}(q) \theta_{4,10}(q) - \theta_{6,10}(q) \theta_{8,10}(q) \Biggr ],
\label{eq:a8}
\end{eqnarray}
and,
\begin{eqnarray}
\lefteqn{\prod_{n=0}^\infty \frac{\left(1\pm q^{n+\half}\right)
{}~\left(1-q^{n+1}\right)~\left(1-q^{5n+5}\right)}
{\left(1+q^{5n+5}\right)~\left(1\mp q^{5n+\frac{5}{2}}\right)}}
\nonumber \\
&=&
\left [ \sum_{n=-\infty}^{+\infty} q^{\frac{5}{2}n^2 +\frac{n}{2}} \right ]
\left [ \sum_{m=-\infty}^{+\infty} (\mp 1)^m q^{\frac{5}{2}m^2 +m} \right ]
\pm q^\half \left [ \sum_{n=-\infty}^{+\infty}
q^{\frac{5}{2}n^2 +\frac{3}{2}n} \right ]
\left [ \sum_{m=-\infty}^{+\infty} (\mp 1)^m q^{\frac{5}{2}m^2 +2m} \right ]
\nonumber \\
&=&
q^{-\frac{5}{40}}
\Biggl (
\left [ \theta_{1,10}(q)+\theta_{9,10}(q) \right ]
{}~\left [ \theta_{2,10}(q)\mp \theta_{8,10} (q)\right ]
\pm
\left [ \theta_{3,10}(q)+\theta_{7,10}(q) \right ]
{}~\left [ \theta_{4,10}(q)\mp \theta_{6,10}(q) \right ] \Biggr ),\nonumber \\
\label{eq:a9}
\end{eqnarray}
respectively.

The product of two theta functions at levels $k$ and $k'$ is
\begin{equation}
\theta_{m,k} (q) ~\theta_{m',k'} (q) = \sum_{\ell = 1}^{k+k'}
\theta_{mk'-m'k+2\ell kk',kk'(k+k')} (q)~\theta_{m+m'+2\ell k,k+k'} (q).
\label{eq:prod}
\end{equation}
In particular, we have,
\begin{eqnarray}
\theta_{2,10}(q) \theta_{4,10}(q) - \theta_{6,10}(q) \theta_{8,10}(q)&=&
\left [\theta_{2,20}(q) - \theta_{18,20}(q) \right]
{}~\left [\theta_{6,20}(q) - \theta_{14,20}(q) \right] \nonumber \\
&=& \eta^2 (q) \chi^{Vir~(4)}_{2,1} (q) ~\chi^{Vir~(4)}_{2,2} (q),
\label{eq:a11}
\end{eqnarray}
and,
\begin{eqnarray}
\lefteqn{\biggl [ \theta_{1,10}(q)+\theta_{9,10}(q) \biggr ]
{}~\biggl  [ \theta_{2,10}(q)\mp \theta_{8,10} (q)\biggr ]
\pm
\biggl  [ \theta_{3,10}(q)+\theta_{7,10}(q) \biggr ]
{}~\biggl  [ \theta_{4,10}(q)\mp \theta_{6,10}(q) \biggr ]}\nonumber \\
&=&
\biggl  [ \theta_{3,20}(q) \mp \theta_{17,20}(q)
\pm  \theta_{7,20}(q)  -\theta_{13,20}(q) \biggr ]
\biggl  [ \theta_{1,20}(q) \mp \theta_{19,20}(q)
\pm  \theta_{11,20}(q) -\theta_{9,20}(q)  \biggr ]
\nonumber \\
&=&
\eta^2 (q)
 \biggl [ \chi^{Vir~(4)}_{1,1} (q) \pm \chi^{Vir~(4)}_{3,1} (q) \biggr ]
{}~\biggl [ \chi^{Vir~(4)}_{1,2}(q)  \pm \chi^{Vir~(4)}_{3,2} (q) \biggr ].
\label{eq:a12}
\end{eqnarray}
We repeatedly used the relation,
\begin{equation}
\theta_{m,20}(q)
= \sum_{p=-\infty}^{+\infty} ~\sum_{\ell=0}^9
q^{20(10p+\ell)^2 + (10p+\ell)m + \frac{m^2}{80} }
= \sum_{\ell=0}^9 \theta_{10m + 400 \ell,2000}(q),
\end{equation}
and the definition of Virasoro characters \eqn{vir}.
Using the product representation of level $m=4$ Virasoro characters once more,
it is very easy to see from \eqn{a11} and \eqn{a12} that the identities
\eqn{id1} hold.

We now proceed to prove the identity \eqn{id2},
\begin{equation}
\chi^{Vir~(4)}_{2,2} (q) = \chi^{Vir~(3)}_{2,2} (q)
{}~\left ( \chi^{Vir~(5)}_{1,1} (q)-\chi^{Vir~(5)}_{4,1} (q) \right ),
\label{eq:id2proof}
\end{equation}
which we rewrite using \eqn{chi3} and \eqn{chi4} and the definition of
$m=5$ Virasoro characters in terms of theta functions as,
\begin{eqnarray}
\prod_{n=0}^\infty \left(1- q^{n+1}\right)
&=&
q^{-\frac{1}{120}}~
\prod_{n=0}^\infty \left(1+ q^{5n+1}\right)~\left(1+ q^{5n+4}\right)
{}~\left(1- q^{5n+1}\right)~\left(1- q^{5n+4}\right)\nonumber \\
&\times &
\left [ \theta_{1,30}(q) + \theta_{29,30}(q)
-  \theta_{11,30}(q)-\theta_{19,30}(q) \right ].
\end{eqnarray}
The infinite product on the RHS can be expressed as a product of
level $k=30$ theta functions after use of the Jacobi triple identity,
\begin{eqnarray}
\lefteqn{q^{-\frac{1}{120}}~
\prod_{n=0}^\infty \left(1+ q^{5n+1}\right)~\left(1+ q^{5n+4}\right)
{}~\left(1- q^{5n+1}\right)~\left(1- q^{5n+4}\right) }\nonumber \\
&&
= q^{-4+\frac{5}{24}} \prod_{n=0}^\infty \left(1- q^{15n+15}\right)^{-6}
\prod_{r=0}^2\left[ \theta^2_{27-10r,30}(q) - \theta^2_{3+10r,30}(q) \right ],
\end{eqnarray}
while the LHS is given by,
\begin{equation}
\prod_{n=0}^\infty \left(1- q^{n+1}\right)
= q^{-4+\frac{5}{24}} \prod_{n=0}^\infty \left(1- q^{15n+15}\right)^{-6}
\prod_{r=0}^6
\left[ \theta_{2r+1,30}(q)- \theta_{29-2r,30} (q) \right ].
\end{equation}
The identity to prove then reduces to,
\begin{eqnarray}
\lefteqn{\prod_{r=0}^2
\left[ \theta_{9+10r,30}(q)- \theta_{21-10r,30} (q) \right ]
{}~\left[ \theta_{5,30}(q)- \theta_{25,30} (q) \right ] }\nonumber \\
&=&
\prod_{r=0}^2
\left[ \theta_{7+10r,30}(q)+ \theta_{23-10r,30} (q) \right ]
{}~\left [ \theta_{1,30}(q) + \theta_{29,30}(q)
-  \theta_{11,30}(q)-\theta_{19,30}(q) \right ],
\end{eqnarray}
which is equivalent to,
\begin{eqnarray}
\lefteqn{ \left [ \theta_{2,15}(q) ~
\theta_{5,15}(q)+\theta_{10,15}(q) ~\theta_{13,15}(q)\right ]
{}~\left [ \theta_{1,15}(q) ~  \theta_{12,15}(q)+\theta_{3,15}(q)
{}~\theta_{14,15}(q)\right ]}\nonumber \\
&=&
\left [ \theta_{3,15}(q) ~  \theta_{10,15}(q)+\theta_{5,15}(q)
{}~\theta_{12,15}(q)\right ]
{}~\left [ \theta_{4,15}(q) ~  \theta_{7,15}(q)+\theta_{8,15}(q)
{}~\theta_{11,15}(q)\right ],
\label{eq:a19}
\end{eqnarray}
as can be seen by using the product of two theta functions at level 30
\eqn{prod}
as well as the standard properties,
\begin{equation}
\theta_{m,60} (q) = \sum_{\ell = 0}^{29} \theta_{30m+3600\ell,54000} (q),
\label{eq:tprop1}
\end{equation}
and,
\begin{equation}
\theta_{m,15} (q) = \theta_{2m,60} (q) + \theta_{60-2m,60} (q).
\label{eq:tprop2}
\end{equation}
This last equality \eqn{a19} can be derived by considering the product,
\begin{equation}
\left[ \theta_{3,30}(q)+ \theta_{27,30} (q) \right ]
{}~\left[ \theta_{7,30}(q)+ \theta_{23,30} (q) \right ]
{}~\left[ \theta_{11,30}(q)+ \theta_{19,30} (q) \right ]
{}~\left[ \theta_{13,30}(q)+ \theta_{17,30} (q) \right ],
\end{equation}
and using \eqn{prod}, \eqn{tprop1} and \eqn{tprop2} by pairing the
first and second factors together (and the third and fourth factors)
and then by pairing the first and third factors together. This completes
the proof of  \eqn{id2proof}.  The other identities of \eqn{id2} can be
proved along similar lines.

Finally we wish to show that the bilinear in $m=5$ Virasoro characters
\eqn{d5} is equal to 1,
\begin{equation}
\epsilon_{ab} \left ( \chi^{Vir~(5)}_{a,1} (q)
+ \chi^{Vir~(5)}_{a,5} (q) \right ) \chi^{Vir~(5)}_{b,3} (q) = 1.
\end{equation}
 From the definition of Virasoro characters in terms of theta functions
\eqn{vir}
and using the same techniques as above, we must prove that,
\begin{equation}
\eta^2(q) = \left [ \theta_{1,15}(q)  - \theta_{11,15} (q)\right ]
{}~\left [ \theta_{2,15}(q)  - \theta_{8,15} (q)\right ]
-\left [ \theta_{4,15}(q)  - \theta_{14,15} (q)\right ]
{}~\left [ \theta_{7,15}(q)  - \theta_{13,15} (q)\right ].
\end{equation}
However, the Euler pentagonal identity \cite{And} gives an expression
of the Dedekind function in terms of level 6 theta functions,
\begin{equation}
\eta(q)  = q^{\frac{1}{24}} \sum_{m=-\infty}^{+\infty}
(-1)^m q^{\frac{3}{2}m^2 + \half m} = \theta_{1,6} (q) - \theta_{5,6}(q),
\end{equation}
and therefore,
\begin{eqnarray}
\eta^2(q) &=& \left [\theta_{1,6} (q) - \theta_{5,6}(q)\right ]^2 \nonumber \\
&=& \theta_{0,3} (q) \theta_{1,3}(q)-\theta_{2,3} (q) \theta_{3,3}(q)\nonumber
\\
&=& \left [\theta_{0,12} (q) + \theta_{12,12}(q)\right ]~\theta_{1,3} (q)
- \left [\theta_{4,12} (q) + \theta_{8,12}(q)\right ]~\theta_{3,3} (q)\nonumber
\\
&=& \left [ \theta_{1,15}(q)  - \theta_{11,15} (q)\right ]
{}~\left [ \theta_{2,15}(q)  - \theta_{8,15} (q)\right ]
-\left [ \theta_{4,15}(q)  - \theta_{14,15} (q)\right ]
{}~\left [ \theta_{7,15}(q)  - \theta_{13,15} (q)\right ],\nonumber \\
\end{eqnarray}
where the product formula \eqn{prod} and \eqn{tprop2} have been used.

\end{document}